\documentclass[a4paper, amsfonts, amssymb, amsmath, reprint, showkeys, nofootinbib, twoside]{revtex4-1}

\usepackage[english]{babel}
\usepackage[utf8]{inputenc}

\usepackage[table,colorinlistoftodos]{xcolor}
\usepackage[color=green!40, prependcaption]{todonotes}
\usepackage{colortbl}
\usepackage{orcidlink}
\usepackage{graphicx}
\DeclareGraphicsExtensions{.pdf,.png,.jpg}
\usepackage{academicons}

\definecolor{orcidlogocol}{HTML}{A6CE39}

\usepackage{amsthm}
\usepackage{mathtools}
\usepackage{physics}
\usepackage{xcolor}
\usepackage{graphicx}
\usepackage[left=23mm,right=13mm,top=35mm,columnsep=15pt]{geometry} 
\usepackage{adjustbox}
\usepackage{placeins}
\usepackage[T1]{fontenc}
\usepackage{lipsum}
\usepackage{csquotes}

\begin{document}
\title{Quasi-Perfect State Transfer in Spin Chains via Parametrization of On-Site Energies}

\author{
    F. Bezaz$^{1,\dagger}$\orcidlink{0009-0008-0356-2742}\footnotemark[1], 
    CC. Nelmes$^{1,\dagger}$\orcidlink{0009-0002-1686-6282}\footnotemark[1], 
    MP. Estarellas$^{1,2}$\orcidlink{0000-0003-2734-9707}, 
    TP. Spiller$^1$\orcidlink{0000-0003-1083-2604}, 
    and I. D'Amico$^1$\orcidlink{0000-0002-4794-1348}
}

\email[Correspondence email:]{c.nelmes@york.ac.uk; \\irene.damico@york.ac.uk}

\affiliation{$^1$Department of Physics, University of York, York, YO10 5DD, United Kingdom}
\affiliation{$^2$Qilimanjaro Quantum Tech, Carrer de Veneçuela, 74, Sant Martí, 08019 Barcelona, Spain}

\date{\today}

\renewcommand{\thefootnote}{\fnsymbol{footnote}}  % Switch to symbol footnotes († is symbol 2)
\footnotetext[2]{These authors contributed equally to this work.}
\begin{abstract}
In recent years, significant progress has been made in the field of state transfer in spin chains, with the aim of achieving perfect state transfer for quantum information processing applications. Previous research has mainly focused on manipulating inter-site couplings within spin chains; here, we investigate in detail the potential of modifying on-site energies to facilitate precise quantum information transfer. Our findings demonstrate that through targeted adjustments to the diagonal elements of the XY Hamiltonian and leveraging a genetic algorithm, quasi-perfect state transfer can be achieved with careful consideration of the system's spectral characteristics. This investigation into on-site energies offers an alternative approach for achieving high-fidelity state transfer, especially in cases where manipulation of inter-site couplings may be impractical. This study thus represents a significant advancement towards unlocking the diverse applications of spin chains within practical quantum information systems.
\end{abstract}


\maketitle

\section{Introduction} \label{sec:outline}
In the field of quantum information processing (QIP), achieving (quasi) perfect state transfer (PST) across spin chains is highly desirable for the development of distributed quantum computation and networking technologies. Using spin chains as a model for prospective quantum wires has been of particular interest. Since the original proposal \cite{Bose_2003}, investigations have included: uniformly coupled chains \cite{Bose_2003,Bose_2007}; PST in modulated chains \cite{Albanese}, the importance of mirroring \cite{Smallvaryj} and the underpinning mathematics \cite{Vinet}; PST in networks \cite{Christandl1,Christandl2} and the underpinning mathematics \cite{Godsil}; PST in ring systems \cite{TJ}; entanglement generation \cite{Yungb,Kay,M.P}; transfer robustness \cite{Stolze2014,M.P}; application to information processing \cite{Yungb,Kay}; the relationship to and realisation of transport in neutral atom \cite{Li} and electronic \cite{N_2004} systems; transfer speed limit \cite{Yunga} and exceeding this limit \cite{Xie}; quasi-perfect state transfer (QPST) \cite{Vinet2,2019} and the underpinning mathematics \cite{Severini}; application of quantum control \cite{Zhang}; transfer of multi-qubit states \cite{Apo}. This paper expands upon existing research by exploring a comparatively much less investigated approach to QPST \cite{Vinet2,2019}: the modification of on-site energies in spin chains. Unlike traditional methods, that focus primarily on adjusting couplings between sites, our study examines the impact of altering the on-site energies of the XY Hamiltonian, whilst maintaining constant inter-site couplings, to facilitate QPST.
Quantum computing \cite{Nielsen-Chuang} promises significant advancements in computational capabilities, in the current NISQ era \cite{Preskill} but particularly for tasks that are beyond the reach of classical computing systems \cite{QS, Madsen2022}. A key challenge in this domain is the preservation of coherence in quantum information systems during transfer processes, necessitating innovative solutions to ensure high-fidelity transfer. Previous research has largely concentrated on the variation and optimization of coupling strengths within spin chains and networks to achieve PST, and by virtue of codifying the appropriate spectra to elicit mirror inversion \cite{Albanese,Christandl1, Christandl2,Smallvaryj,Kay}. However, the potential of on-site energy adjustments to attain PST has been explored much less.
Though limited, research has been undertaken aiming to exploit optimization techniques, such as genetic algorithms \cite{ES,LC}, for superdense coding \cite{Domínguez-Serna_2015} and to uncover spin chain configurations which yield high-fidelity transfer \cite{Luke}. The aforementioned works, which have predominantly concentrated on coupling configurations within linear systems and on interactions involving coupling and multiple excitations within networks, provide the impetus for our approach. Relevant literature has established that PST is  unattainable for uniformly coupled graphs/chains, with $N \geq 4$ \cite{Christandl1,Godsil, Severini, 2019}, when the on-site energy contributions are set to a site-independent constant. Here, we demonstrate that QPST may still be attained within a range of uniformly coupled chains, only requiring appropriate tuning of the on-site potentials.

The focus on parameterizing on-site energies unveils additional pathways to achieve high-fidelity solutions and presents new flexibility to designing quantum information transmission systems, useful for hardware where engineering of coupling between qubits may be difficult or limited. This contribution aligns with ongoing endeavours to materialize quantum computing applications. To our knowledge, there has been limited research on this type of engineering of spin chains. In fact, previous research has discussed solutions for a single chain of a specific-length and/or the tuning of a global harmonic potential for a ferromagnetic chain \cite{Shi} or an optical lattice system \cite{G-B}, or the creation of dual-site "barrier" potentials for high transmission probability \cite{Lorenzo}. In contrast, the results presented here are systematic and offer a number of prescriptions for varied (odd and even) chain lengths with distinct "families" of solutions based on the spectral properties of the system, whilst still respecting the homogeneity of the spin-spin coupling scheme. 
\\
  %I believe leaving the sections in separate files is more organized, change it if you desire 
\section{The Spin Chain Model and state transfer}\label{sec:Hamiltonian}
The XY Hamiltonian, a widely-applicable, physically relevant model which captures the interaction dynamics between nearest-neighbour spins in a one-dimensional $N$-site chain, is given by
\begin{align}
\Hat{H}_{XY} = & \sum_{i=1}^{N-1} J_{i,i+1}\left(|1\rangle \langle0|_{i} \otimes |0\rangle \langle1|_{i+1} \right. \nonumber \\
         & \left. + |0\rangle \langle1|_{i} \otimes |1\rangle \langle0|_{i+1}\right) + \sum_{i=1}^{N} \epsilon_{i} |1\rangle \langle1|_{i}.
         \label{Hami}
\end{align}
Here, $J_{i,i+1}$ denotes the coupling strength between the spins at sites $i$ and $i+1$, and $\varepsilon_i$ represents the on-site energy at site $i$. By adjusting $\varepsilon_i$, we explore various configurations to achieve QPST. Using this time-independent Hamiltonian, we observe the evolution of the initialization via the natural dynamics of the system, described by the Schr\"{o}dinger equation. In the work presented here we restrict ourselves to the single excitation subspace of the chain, which has a basis of $N$ states.
As a specific example of the general state-mirroring dynamics, throughout our analysis we denote $|\psi\rangle_A$ = $|100...00 \rangle$ as the initial state or input and $|\psi\rangle_B$ = $|000...01 \rangle$ as our target state or desired output. This translates to an initial excitation at the start of the chain and effectively transferring the excitation to the opposite end at some predictable time $t_m$, which may be referred to as the "mirroring time", i.e.
\begin{eqnarray}
|\psi\rangle_A &=& |100\ldots00 \rangle \stackrel{t_m}{\rightarrow} |\psi\rangle_B = |000\ldots01 \rangle.
\label{trans}
\end{eqnarray}
In quantum information systems, nearing PST necessitates the propagation of a quantum state from site A to site B with fidelity approaching unity. For pure states, fidelity with respect to a state $|\psi_i\rangle$ is expressed as $F(|\psi_i\rangle, |\psi_j\rangle) = |\langle \psi_i | \psi_j \rangle|^2$. The temporal evolution of the state $|\psi_A\rangle$ to $|\psi_B\rangle$ across the chain, governed by the Hamiltonian $H$, yields fidelity $F(t) = |\langle \psi_B | e^{-iHt/\hbar} | \psi(0)_A \rangle|^2$. Fidelity approaching unity, underscoring the system's efficacy in quantum information preservation during transfer, is pivotal for optimal quantum communication protocols. Established literature identifies a coupling scheme enabling fidelities of unity, as demonstrated in \cite{Christandl1, N_2004}. This scheme employs a symmetric hopping configuration about the chain's midpoint, ensuring uniform energy spacing across the spectrum for perfect state transfer. The coupling constant between adjacent sites is given by
\begin{eqnarray}
   J_{i,i+1} = J_0\sqrt{(N-i)i}, 
   \label{christa}
\end{eqnarray}
where $J_0$ sets the scale for the coupling strengths, $N$ represents the total number of sites, and $i=1,N$ indexes sites from the chain's beginning. This protocol underlines the important example of equidistant energy levels in facilitating optimal quantum state transfer. The coupling profile described by Eq. (\ref{christa}) requires that the maximum coupling strength, \(J_{\text{max}}\), be located at the centre of the chain, with its relative magnitude compared to the terminal site couplings being proportional to \(N\). In the case of a homogeneous coupling scheme, we set \(J_{\text{max}}=1\), along with the rest of the couplings. $J_{\text{max}}$ will serve as the referential energy unit throughout our analysis, as discussed in Sec. \ref{sec:results}. 

Achieving PST necessitates that the system's Hamiltonian exhibits a "mirror" symmetry with respect to the chain's mid-point \cite{Albanese,Bose_2007, Christandl2,Kay}. This is imposed by the Hamiltonian Eq. (\ref{Hami}) and the mirror matrix  
\begin{equation*}
M = \left( \begin{array}{ccccc}
  0 & 0 & \cdots & 0 & 1 \\
  0 & 0 & \cdots & 1 & 0 \\
  \vdots & \vdots & \ddots & \vdots & \vdots \\
  0 & 1 & \cdots & 0 & 0 \\
  1 & 0 & \cdots & 0 & 0 \\
\end{array} \right)_{N \times N},
\end{equation*}

commuting ( $[H,M] = 0$ ), and therefore enforces
\begin{align}
    \varepsilon_{i} &= \varepsilon_{N-i+1}, \label{ep} \\
    J_{i,i+1} &= J_{N-i, N-i+1},
\end{align}
for all $i \in (1, 2, 3,.., N-1)$. For the information transfer (Eq. (\ref{trans})) to occur perfectly, we require the time evolution operator $U(t,0)$ at time $t_m$ 
\begin{equation*}
\Hat{U}(t_m,0)|\psi_A\rangle = e^{-iH t_m}|\psi_A\rangle = \Hat{M}|\psi_A\rangle = |\psi_B\rangle,
\end{equation*}
where $\hbar = 1$. It may be shown from the above conditions and the periodicity  of the perfect mirroring process -- $2 \cdot t_m$ brings us back to $|\psi_A\rangle$ -- that the eigenvalues capable of delivering fidelities equal to unity are restricted by 
\begin{equation*}
\frac{E_n - E_m}{E_n' - E_m'} \in \mathbb{Q},
\end{equation*}
where $E_n$ is defined as the $n^{th}$ eigenvalue of our Hamiltonian, or equivalently
\begin{equation}
    \Delta E_{n,n+1} = \frac{\pi}{t_m} O_{n,n+1},
    \label{eig}
\end{equation}
where $O_{n,n+1}$ is a pair-dependent odd integer for each sequential pair of eigenvalues \cite{Kay}. This criterion underscores the system's periodicity as a fundamental requirement for PST, supported by rational ratios of eigenvalue differences \cite{Christandl1, Godsil}. An equally-spaced spectrum forms a simple example of this.

\section{Tuning $\varepsilon_i$ for PST}\label{sec:epsilon}
\subsection{p-factor}
It can be demonstrated that perfect state transfer in an XY spin chain, is only achievable up to a 3-site chain when all on-site energies and couplings are respectively equal \cite{Christandl1,Godsil,Severini}. As a result, there have been investigations into the foundations of this observed phenomenon and the codifying of a separate branch of high-fidelity solutions or 'pretty good state transfer' (PGST) for $N > 3$ \cite{ Godsil,Severini, 2019, Vinet2}. In this work, we propose that by appropriately modulating the on-site energies in an otherwise uniformly coupled spin chain, quasi-perfect state transfer can be achieved.

We start with a mirror-symmetric linear chain and consider a spectrum of eigenvalues $E_n$, where all eigenvalues are equally spaced, apart from the highest two, which exhibit a separation one-third of that of the rest of the spectrum \\
\[
 E_n \in \left\{ 5\alpha/3, \alpha, -\alpha, -3\alpha \right\},
\]
with \( \alpha \) being a constant. From Eq. (\ref{eig}) this spectrum translates to 
\begin{equation*}
 O_{n,n+1} = \left\{
\begin{array}{ll}
3 & \textnormal{for} \quad n = 1, 2, \ldots, N-2 \\
1 & \textnormal{for} \quad n = N-1
\end{array} \right.
\end{equation*}
 yielding a non-equidistant upper eigenvalue "pinch" spacing, which is distinct from the previous spectra investigated for PST \cite{Albanese,Christandl1,Christandl2}. It may be shown (see Appendix.) through the eigenvalues fulfilling the necessary requirements Eq. (\ref{eig}), that a state evolved by the corresponding Hamiltonian will mirror invert perfectly at time \( t_m = \frac{3\pi}{2\alpha}\), and therefore the appropriate configurations of the on-site energies and couplings only need to be determined.
  We may now denote the pinch spectra with respect to a constant, where the separation ratio is an odd integer denoted as $p$ and describes the relative strength of the pinch 
\[
\frac{1}{p} = \frac{d'}{\frac{1}{N} \sum_{n=1}^{N} d_n},
\]
where $d_n = |E_n - E_{n+1}|$ and $d'$, the difference between the largest two eigenvalues. The time for perfect state mirroring to occur becomes
\begin{equation}
t_m = \frac{p \pi}{2\alpha},
\label{time}
\end{equation}
and we should therefore have to wait incrementally longer for each odd integer increase of $p$ for the initial excitation to be observed on the opposite end of the chain.  
\subsection{3-site chain}

\begin{figure}[h!] 
    \centering\includegraphics[width=1.0\linewidth]{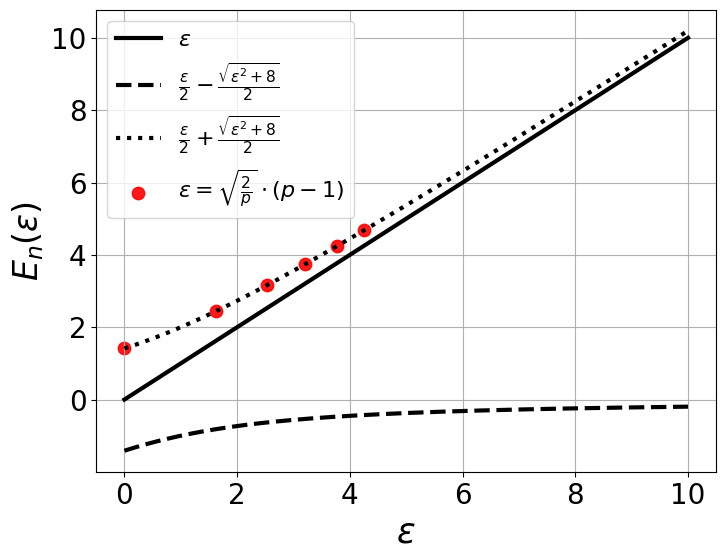}
 % Adjust the value as needed to cut off a little bit of the bottom
    \caption{Eigenvalue equations for $N = 3$, (with central on-site energy set to zero) plotted as a function of $\varepsilon$, the two outer on-site energies, with the corresponding values of the on-site energies from the numerical results. The highlighted points correspond to the values of epsilon such that the spacing between the top two eigenvalues and the lowest two eigenvalues is in the ratio $\frac{1}{p}$; for $p = 1, 3, 5, 7, 9, 11$.}
    \label{analytics+numer}
\end{figure}

For a clearer analytical discussion we may begin with an $N = 3$-XY spin chain, setting $\varepsilon_1 = \varepsilon_3 = \varepsilon$, $\varepsilon_2 = 0$ and all $J_{n,n+1} = 1$ for simplicity. Using Eq. (\ref{Hami}) we may derive the following Hamiltonian matrix
\begin{equation}
H_{XY_{(N=3)}} = \left( \begin{array}{ccc}
\varepsilon & 1 & 0 \\
1 & 0 & 1 \\
0 & 1 & \varepsilon 
\end{array} \right).
\label{N3Hami}
\end{equation}
The eigenvalues for Eq. (\ref{N3Hami}) are $E_1 = \varepsilon$ and $E_{2,3} = \frac{\varepsilon}{2}\pm\frac{\sqrt{\varepsilon^2+8}}{2}$ and through further algebra, taking the eigenvalue spacing between the highest eigenvalues to be $\frac{1}{p}$ of the lowest, the expression relating the $p$-factor to the value of \( \varepsilon \) is found to be 
\begin{equation}
\varepsilon = \sqrt{\frac{2}{p}}\cdot(p-1).
\label{eps}
\end{equation}
The values of $\varepsilon$ which correspond to odd integer values of $p$ from 1 to 11, from Eq. (\ref{eps}), may be seen as the highlighted (red) points within Fig. (\ref{analytics+numer}). Clearly, this spectrum arises when the strengths of $\varepsilon_i$ are minimal at the midpoint of the chain, and maximal at the ends, without any inter-spin coupling modulation. For larger $N$, further tuning efforts will then be directed towards optimizing on-site energy values,  guided by an auxiliary search for configurations whose spectra match the pre-selected $p$-values.
\section{Genetic Algorithm: Set-up and Execution}\label{sec:method} \renewcommand{\thefootnote}{\arabic{footnote}}  % Switch back to numbered footnotes
\setcounter{footnote}{1} 
\subsection{Fitness}
The problem of determining an on-site energy profile which maximizes the transfer fidelity across an $N$-site chain can be effectively addressed through evolutionary computation, specifically utilizing a genetic algorithm \cite{ES,LC,Luke}. This algorithmic process commences with an initial population of potential solutions, which are then subjected to a Darwinian selection process based on their performances, as assessed by a predefined fitness function. The fitness function effectively guides the genetic algorithm towards a solution which outperforms others with regard to a specific set of traits. A mutation process—either fixed or dynamically varying—then either expands or narrows the exploration within the search space, balancing the trade-off between global exploration and convergence to local optima.

To search for high-fidelity configurations forming from a desired spectrum (tuned by a selected $p$-value), the fitness function was defined as follows 
\begin{equation}
f(F_{\max},\upsilon; Q,p,\sigma_{E}) = \aleph [(A\cdot F_{\max}) - (B\cdot\upsilon)],
\label{ff}
\end{equation}
where $F_{max}$ is the maximum fidelity score achieved by the configuration within the time frame of evaluation, $\upsilon$ is the cumulative penalty associated with the desired spectrum 
\begin{equation}
\upsilon = |Q - (\frac{1}{p})| + (\sigma_{E_n}),
\label{ups}
\end{equation}
formed by a $Q$-factor
\begin{equation*}
 Q = \frac{(\Delta E_{N-1,N})^{N-2}}{\prod_{n=1}^{N-2}(\Delta E_{n,n+1})},    
\end{equation*}
and the standard deviation $\sigma_{E_n}$ of the eigenvalue spacings, apart from the highest two. Within Eq. (\ref{ff}), $A$ and $B$ are adjustable scaling factors, giving relative weights between the fidelity and spectral penalties, whilst $\aleph$ is an appropriate normalization
\begin{equation*}
\aleph = {(A \cdot F_{\text{max}} + B \cdot \upsilon)}^{-1},
\end{equation*}
so that the fitness is scaled by the combination of the largest contributions of the penalties and the fitness values falls within the range $[0-1]$. It becomes clear that if $A$ is chosen such that $A \gg B$, the genetic algorithm will find configurations which lead to purely higher-fidelity configurations with little consideration for the spacing between eigenvalues within the spectrum.  The reverse is true when $A \ll B $, where the genetic algorithm will search for configurations which have the most desirable spectra and fidelity is a lesser consideration. Particularly for smaller chain lengths (such as $N = 3,4$), $A$ was selected to be of the order of $B$ ($A \approx B$), to give approximately equal weight to the fidelity and spectra, whilst for higher order $N$-site chains, there was a larger weight placed on the fidelity ($A > B$).\footnotemark[2] \footnotetext{There is an expectation of a relative saturation of very high fidelity solutions within smaller $N$-site chains \cite{Bose_2007}.}
\subsection{Algorithm Execution}
\begin{figure}[h!]
    \centering
    \includegraphics[width=0.85\linewidth]{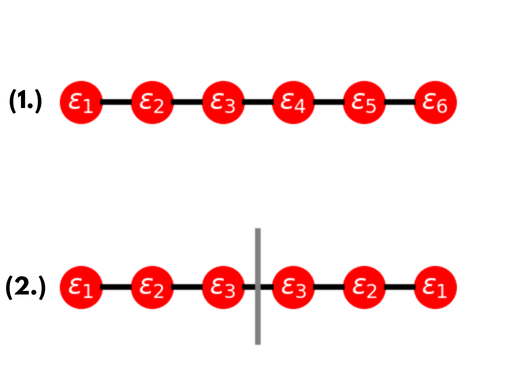}
    \caption{The mutation process illustrated for a 6-site chain. The initial randomized chain (1.) is reflected about its centre (2.) from the left, to enforce mirror symmetry of the on-site energies throughout the execution of the algorithm. The solid black lines connecting the sites imply homogeneous coupling $J_{i,i+1} = 1$.}
    \label{mir}
\end{figure}

The genetic algorithm begins with the selection of an individual created with a randomized onsite Hamiltonian configuration and equal hopping $J_{i,i+1} = 1$, from a population of the desired form Eq. (\ref{Hami}). A subsequent mutation function takes the  individual from the population and uniformly alters the diagonal of the individual by random values within a specified range (1-10 throughout all of the iterations discussed here). It had been decided that in the interest of preserving mirror symmetry about the centre of the chain Eq. (\ref{ep}), as well as greatly narrowing the search space and subsequently the computational taxation required, that the algorithm should only consider mirror symmetric individuals. This is enforced throughout the mutation process via the mutated offspring being mirrored about its centre Fig. \ref{mir}, and then subsequently passed onto the fitness evaluation within the genetic algorithm. 

The natural dynamics of the system are then observed, recording the maximum fidelity attained within the allotted time window, the spacing of the energy spectrum, and the standard deviation of the spacing between the eigenvalues. The information about the individual is then indexed within the genetic algorithm, and its fitness score assessed via the specification defined by the fitness function Eq. (\ref{ff}). The crossover operation to propagate information to the next generation of prospective solutions (offspring) then follows and is in-line with previous research where between selected parents, an evenly-distributed amount of genetic information are exchanged to serve as the genetic material for their children \cite{Luke}. This exchange between the previous generation occurs with equal probability so each parent has an opportunity to pass on 50\% of their respective encoding. Following the crossover stage, the offspring are mutated according to an adjustable, generation-dependent mutation function
\begin{equation}
\mu(g) = \mu_i - g \cdot \frac{\mu_i - \mu_f}{G},
\label{mug}
\end{equation}
where $\mu_i$ and $\mu_f$ are the pre-selected initial and final mutation values, respectively. This function is scaled by the total number of generations $G$ and the current generation $g$, ranging from 0 to $G$.
 Iterations of these mutated individuals, over sufficiently large generations, are selected with increasing fitness scores until the fittest "Darwinian individual" is returned once the algorithm successfully terminates. 
The parameters for generations, initial population size, and mutation rate (including the evolution of the mutation rate) were selected to mirror the methodology employed in prior research on adaptive quantum device design and are shown in Table.\ref{tab:GAparams} \cite{Luke}.  
\begin{table}[h!]
\centering
\small
\begin{tabular}{|l|c|c|}
\hline
\rowcolor{gray!30}
\textbf{Generations} & \(\mathbf{{Population-Size}}\) & \(\mathbf{Mutation-Rate\footnote{Initial mutation rate starts at 20\% and is set to decrease as a function of the generations.}}\)  \\ \hline
\hspace{0.6cm} 200 & 1024 & ${20\%}$\\ \hline
\end{tabular}
\caption{The optimization parameters employed for all of the data presented within Sec.\ref{sec:results}. Note that, though the initial mutation rate is set to 20\%, it was decreased ($\mu_i > \mu_f$), as a function of the number of generations (Eq. (\ref{mug})) to increase the exploration of the local optima.}
\label{tab:GAparams}
\end{table}
The primary alterations from the methodology in the previous paper comes predominantly through the aforementioned mirroring-mutation process (See Fig.\ref{mir}) as well as the new fitness function. Once the parameters are set within the fitness function, $N$ may be adjusted through the scaling of the associated $N \times N$ Hamiltonian matrix, along with the size of the initial (and therefore also the target) state Eq. (\ref{trans}). The value of $p$ was increased in even-integer steps (to the next odd integer) within Eq. (\ref{ups}) to further produce solutions for very high-fidelity transfer with the desired spectral characteristics.

\section{Results}\label{sec:results}
\begin{figure}[h!]
    \centering
    \begin{minipage}{0.8\linewidth}
        \centering
        \hspace{-1cm} % Adjust the amount of horizontal space to shift the plots to the left
        \includegraphics[width=\linewidth]{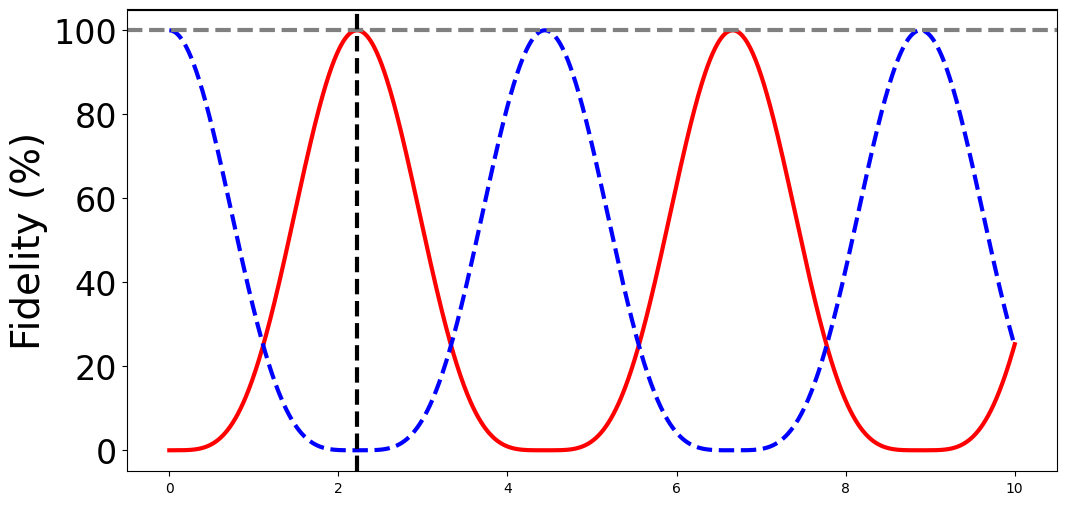}
    \end{minipage}
    
    \vspace{-0.25cm} % Adjust vertical spacing between figures if needed
    
    \begin{minipage}{0.8\linewidth}
        \centering
        \hspace{-1cm} 
        \includegraphics[width=\linewidth]{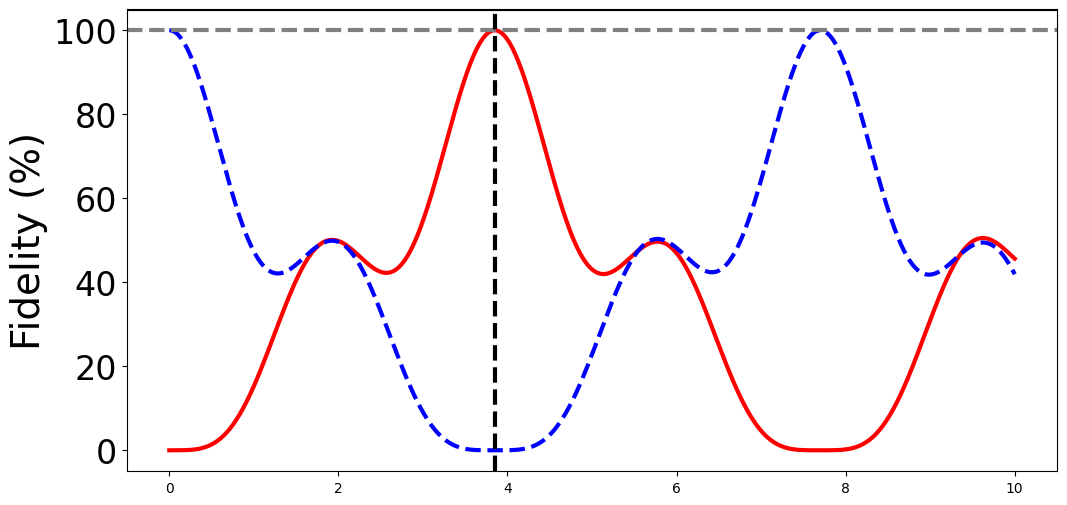}
    \end{minipage}
    
    \vspace{-0.25cm} 
    
    \begin{minipage}{0.8\linewidth}
        \centering
        \hspace{-1cm} % Adjust the amount of horizontal space to shift the plots to the left
        \includegraphics[width=\linewidth]{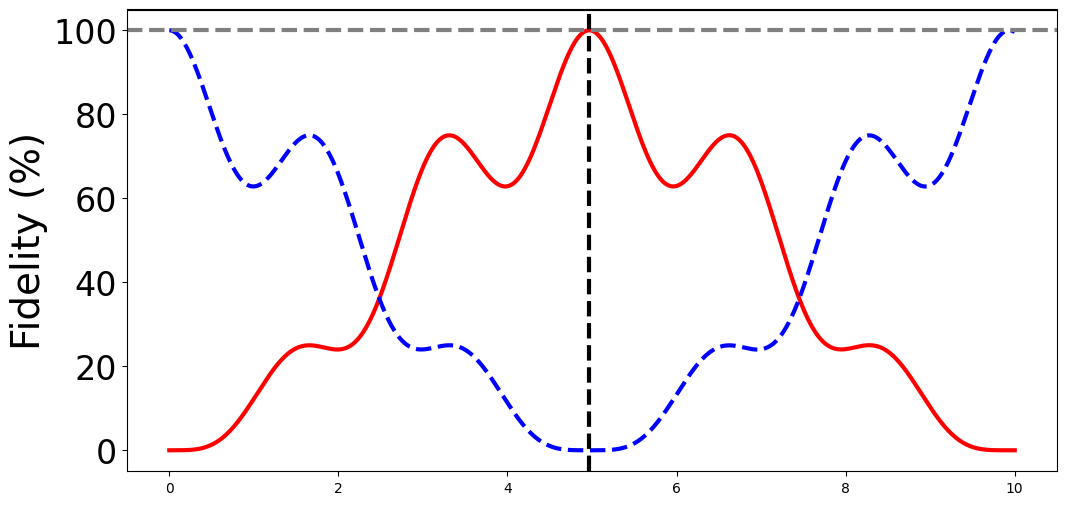}
    \end{minipage}
    
    \vspace{-0.25cm} % Adjust vertical spacing between figures if needed
    
    \begin{minipage}{0.8\linewidth}
        \centering
        \hspace{-1cm} 
        \includegraphics[width=\linewidth]{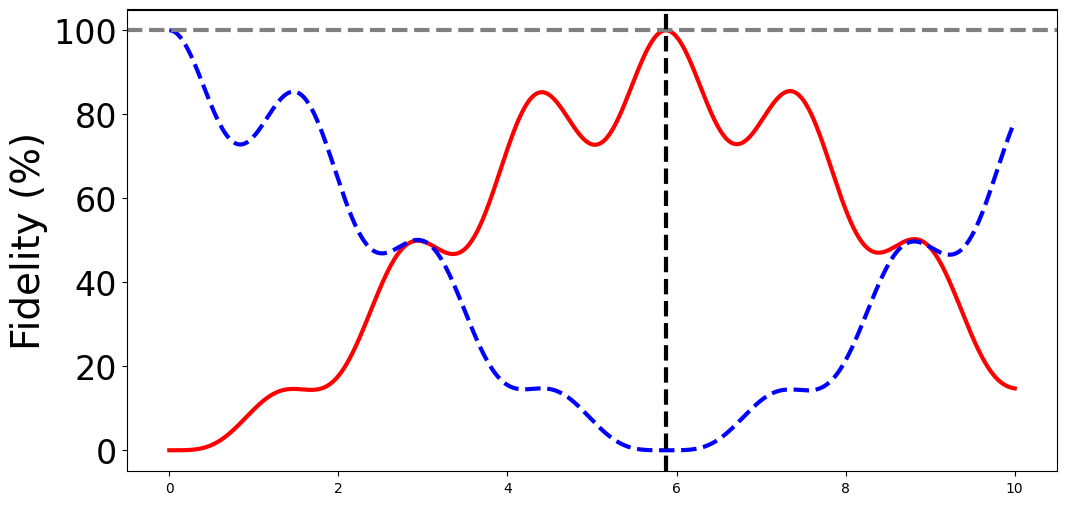}
    \end{minipage}
    
    \vspace{-0.25cm} 
    
    \begin{minipage}{0.8\linewidth}
        \centering
        \hspace{-1cm}
        \includegraphics[width=\linewidth]{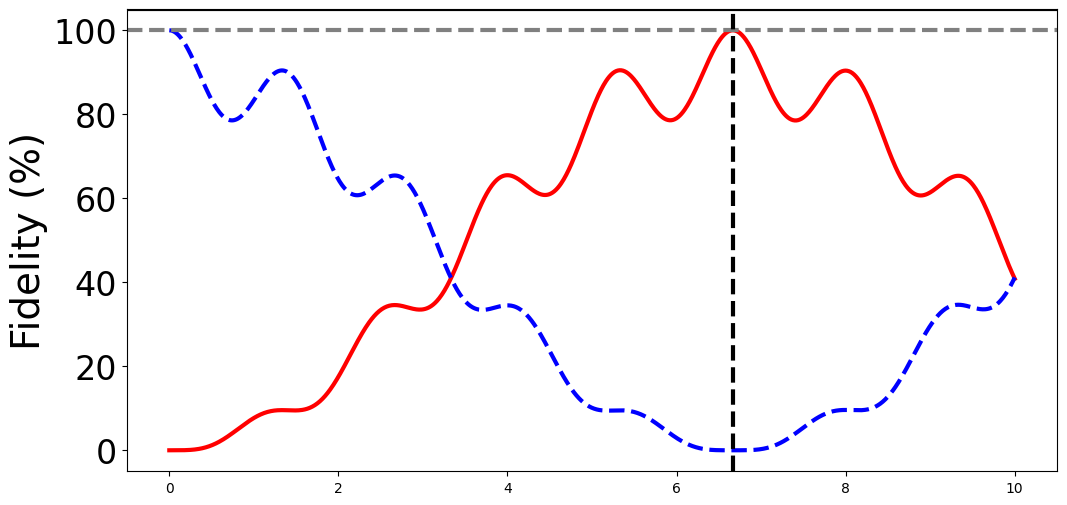}
    \end{minipage}
    
    \vspace{-0.25cm} % Adjust vertical spacing between figures if needed
    
    \begin{minipage}{0.8\linewidth}
        \centering
        \hspace{-1cm} 
        \includegraphics[width=\linewidth]{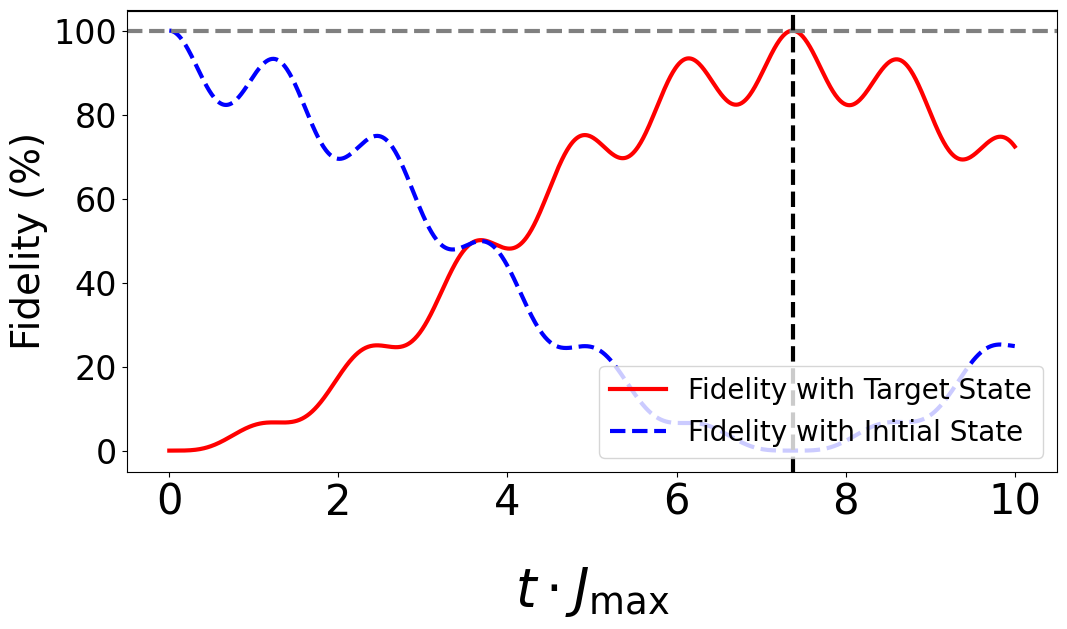}
    \end{minipage}

     \caption{Fidelity with target state (red solid curve) and Fidelity with initial state (blue dashed curve) versus re-scaled time $t \cdot J_{max}$ for $p$ = 1 (top) to 11 (bottom), for $N$=3-site chain. The dashed (grey) horizontal line provides a reference line along fidelity value of 100\% and the dashed (black) vertical line indicates first instance of perfect state transfer.}
    \label{Dynamics}
\end{figure}

\begin{figure}[h!]
    \centering
    \includegraphics[width=0.95\linewidth]{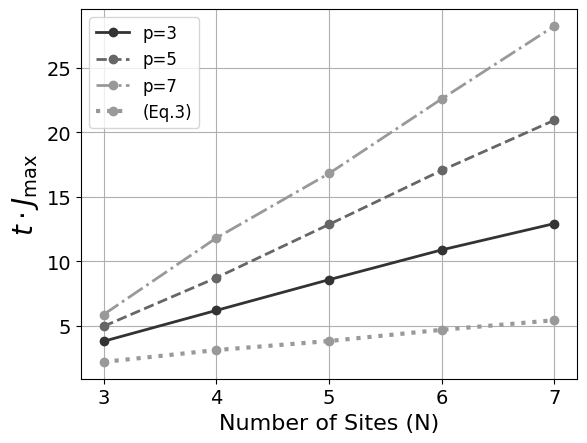}
    \caption{Comparison of times (in units of $t \cdot J_{max}$) to acquire QPST (PST for $N$ = 3) via the on-site energy parametrization, and the previously known coupling scheme (Eq. (\ref{christa})) with no on-site energy variation \cite{Christandl1, N_2004}.}
    \label{families}
\end{figure}

In the following section we outline the results from the genetic algorithm designed for on-site energy parametrization. Any reference to time within the figures shown, proceeds through the use of natural units ($\hbar = 1$), consequently making $J_{max}$ the characteristic energy/inverse time scale. Beginning with a 3-site chain, as a test,  the genetic algorithm was able to obtain the predicted relative values of the on-site energies for odd $p = 3 - 11$. Consequently, the highlighted points within Fig.  \ref{analytics+numer} display perfect correspondence between analytical foundation and numerical findings (Eq. (\ref{eps})). Furthermore, the dynamics for each of the $p$-solutions for
$N = 3$ are shown in Fig.  \ref{Dynamics}, where the increase in time required for successful state transfer is in accordance with Eq. (\ref{time}) ($t_{QPST}> t_{PST}$).
The distinct shapes associated with the fidelity between the evolved state and the target (or the initial) state over time, are formed by the number of "attempts" (we may denote them as $a$) required for the state to mirror invert completely
\begin{equation*}    
a = \frac{(p-1)}{2}.
\end{equation*}
Attempts are characterized by maxima with $F<1$.
Clearly for $p = 1$ no previous attempts are required, making this the most time-efficient approach to PST. The next best option is waiting proportionally to $p$, for the next opportunity. Extending the application of the genetic algorithm to a larger number of $N$-sites, we discovered a set of optimal solutions characterized by specific \( p \) values (3, 5, and 7) across various chain lengths (\( N = 3 \) to 7), as depicted in Fig.  \ref{families}.
The \( p \) solutions for different chain lengths can be grouped into distinct families, with respect to the time required for PST (for \( N = 3 \)) and QPST (for \( N = 4 \) to 7). Each individual \( p \)-family  exhibits an approximately linear dependence on the chain length \( N \), with a gradient that increases with increasing $p$.

\begin{figure}[h!]
    \centering
    
    \begin{minipage}{1.0\linewidth}
        \centering
        \includegraphics[width=1.0\linewidth]{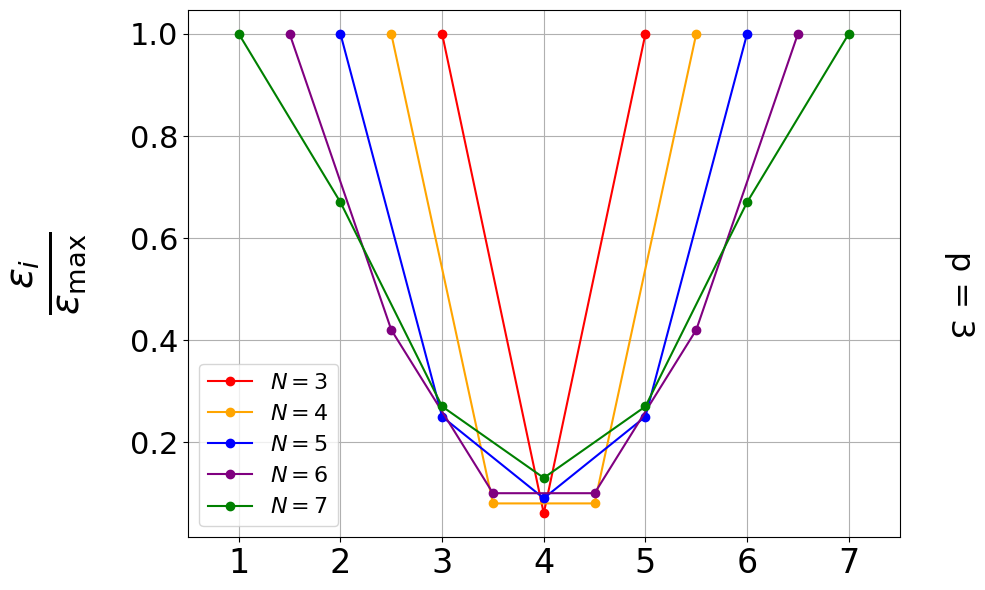} % Adjust the width value as needed
    \end{minipage}
    
    \vspace{-0.45cm} % Adjust vertical spacing between figures if needed
    
    \begin{minipage}{1.0\linewidth}
        \centering
        \includegraphics[width=1.0\linewidth]{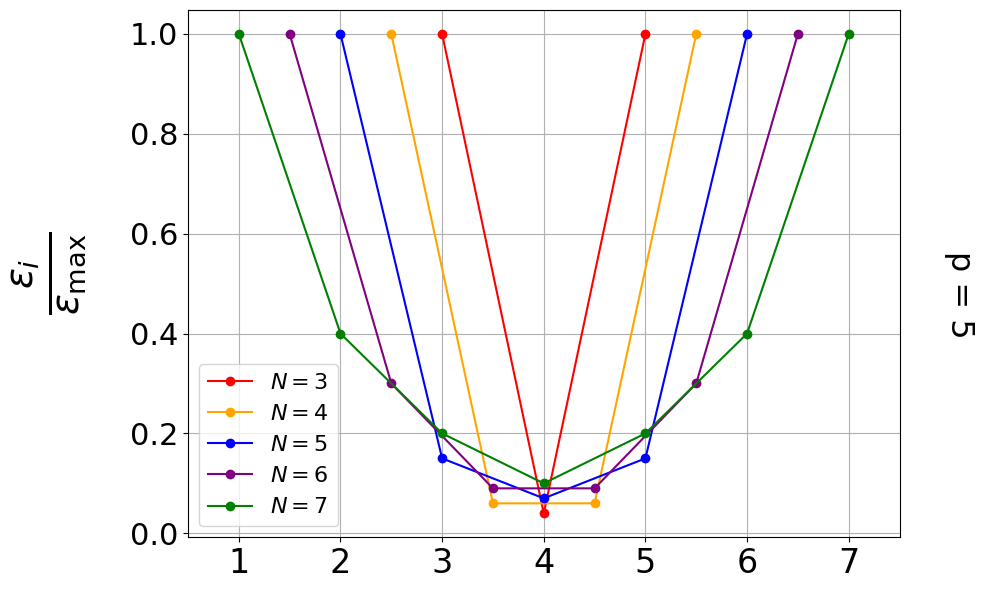} % Adjust the width value as needed
    \end{minipage}
    
    \vspace{-0.41cm} % Adjust vertical spacing between figures if needed
    
    \begin{minipage}{1.0\linewidth}
        \centering
        \includegraphics[width=1.0\linewidth]{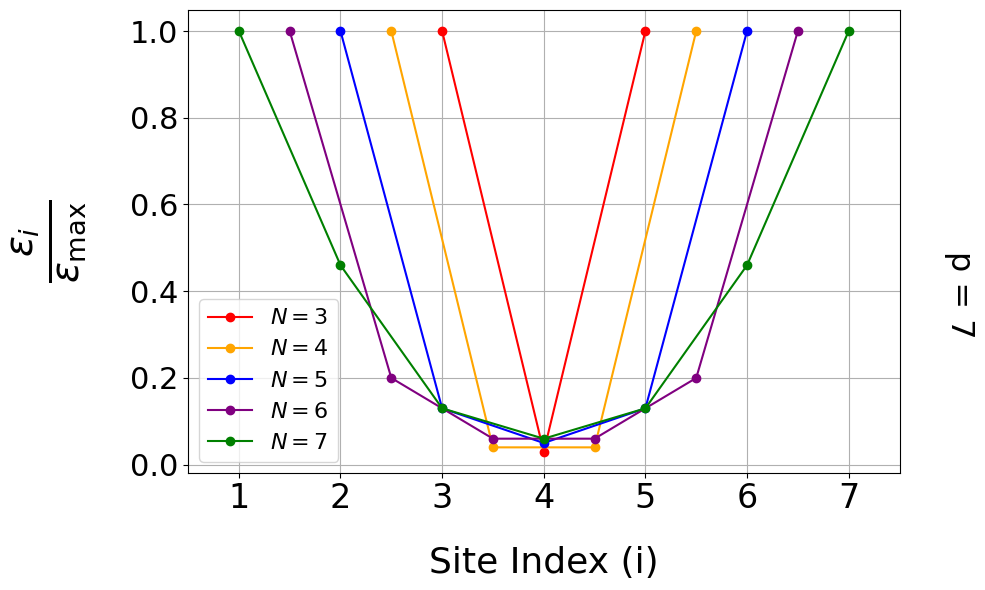} % Adjust the width value as needed
    \end{minipage}
    
    \caption{On-site configuration trends for $N = 1 - 7$ spin chains, considering p-values of 3, 5, and 7 for the solution families, with the x-axis being the site index along the chain. Each plotted point is shifted for $N < 7$ along the x-axis by $(7 - N) - \frac{1}{2}$, aligning each plot's central site/mid-bond center with the midpoint of $N = 7$. This adjustment enables a more convenient comparison of the relative strengths of outer-onsite energies versus central site potentials across different chain lengths and solution families.}
    \label{on}
\end{figure}

\begin{figure}[h!]
    \centering
    \includegraphics[width=0.9\linewidth]{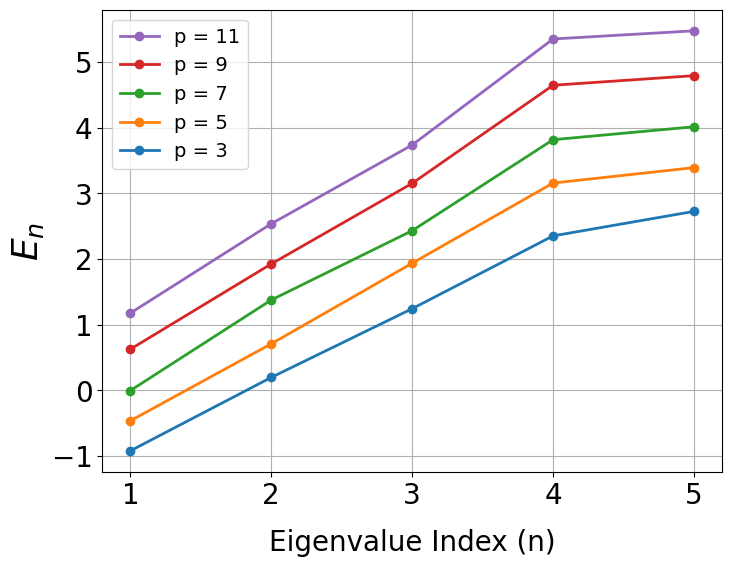}
    \caption{Spectra diagram for $N = 5$; $p = (3 - 11)$, where $E_n$ are in units of $J_{max}$. The plotted eigenvalues were spaced by insisting $\varepsilon_i+i = \varepsilon_i $, so full spectrum for each individual $p$-solution has enough relative spacing to be easily distinguished. The lowest and highest fidelities achieved from the configurations associated to these spectra was 95.1\% and 99.9\%, from p = 11 and 3 respectively.}
    \label{spectra}
\end{figure}

The on-site energy configuration trends  for these different $p$ solutions with respect to the chain's sitesare displayed in Fig. \ref{on}. The trends showcase an approximate parabolicity of the on-site energies about the centre of the chain, similar to the analytical results of the $N$ = 3-site chain, with the inverted triangular configuration (Eq. (\ref{N3Hami})). Furthermore, with few exceptions, the degree of steepness of the descent from the two outer on-site energies towards the middle becomes larger with increasing order of $p$. The general shape and structure of the spectra associated with various strengths of pinching between the highest and second highest eigenvalue can be observed in Fig. \ref{spectra}, where these two eigenvalues tend towards degeneracy as \( p \) increases. The genetic algorithm was also able to find other $p$-solutions for $N > 7$, such as $N = 8$ (discussed in Sec.\ref{prevprot}) and $N = 9$ (not shown), but it becomes progressively more difficult to systematically extend the complete range of families for larger $N$, of the kind exhibited in Fig. \ref{families}. Behaviour of larger $N$ chains is clearly of interest. This is under further investigation and results will be reported in a future paper. 

\subsection{Robustness}
\begin{figure}
    \centering
    \includegraphics[width=1.0\linewidth]{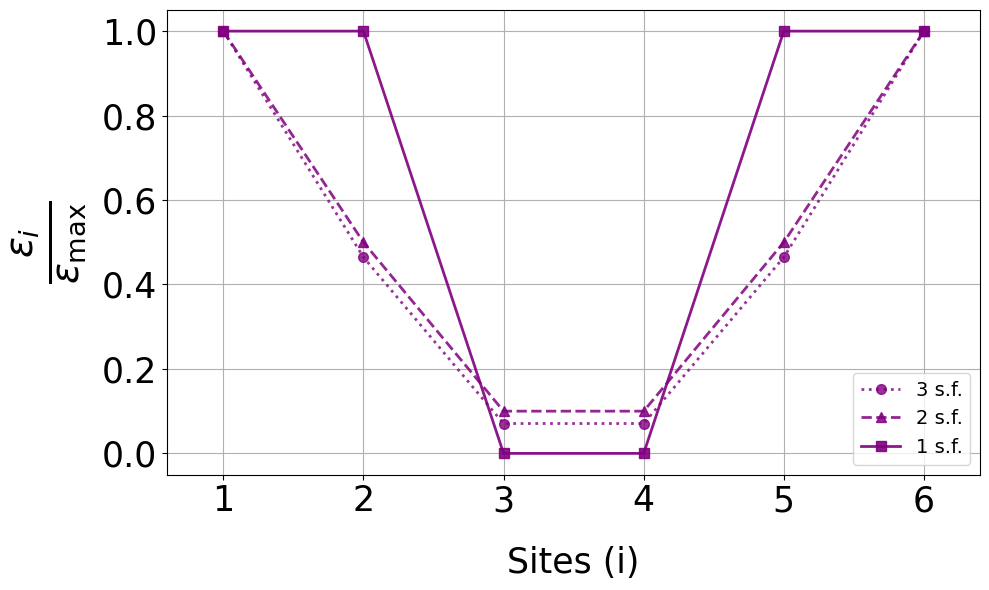}
    \caption{On-site configuration trend for $N = 6$; $p = 3$ chain with decreasing levels of precision. When the values of $\varepsilon_i$ is rounded to 1 significant figure (solid line-square marker), $97.1\%$ fidelity is still achieved (Table \ref{table:efficiency}). The plot for 4 s.f. was omitted as it essentially overlaps with the one with  3 s.f.}
    \label{robust}
\end{figure}

\begin{figure}
    \centering
    \includegraphics[width=1\linewidth]{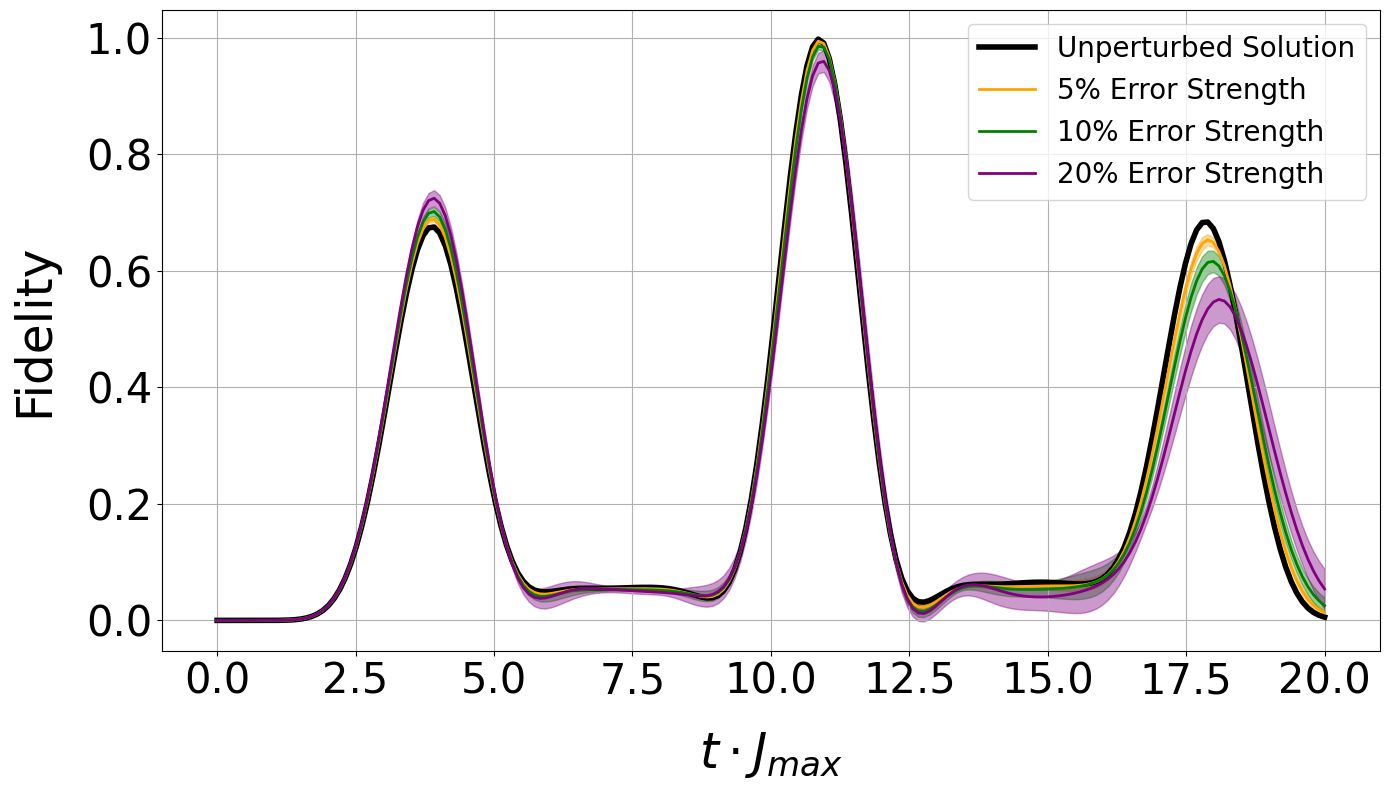}
    \caption{Time evolution of the mean transfer fidelity for the original solution associated with $N=6; p =3$ under varying strengths of stochastic, asymmetric perturbation.
        The plot illustrates the average transfer fidelity as a function of time for three different perturbation levels: $\xi$ = 0.05, 0.1, and 0.2. 
        The maximum mean fidelities of 99.72$ \pm $0.01\%, 99.28$ \pm $0.05\%, and 97.78$ \pm  $0.14\% were achieved for 5\%, 10\%, and 20\% perturbations, respectively.
        Shaded, colored areas represent the standard deviation for each 1,000 simulation sample, 
        highlighting the variability and robustness of the quantum state transfer process under different levels of asymmetric perturbations.}
    \label{Error}
\end{figure}

The fidelities associated with the on-site energy trends displayed in Fig. \ref{on} may be found in Table \ref{table:efficiency}. This table also includes effect of static noise, captured by different degrees of precision required for the values of $\varepsilon_i$ to enable QPST. Notably, even when rounded to just one significant figure, all of the fidelity values displayed in Table~\ref{table:efficiency} are above 81\%, comfortably exceeding the $66.6\overline{6}\%$ threshold, which is recognized as the upper limit for classically achievable means of communication \cite{Bose_2007,Hor}. This demonstrates a high level of robustness to imprecision, as it still allows for very high fidelity transfer, even when each on-site energy value is rounded to 1 s.f. and the corresponding on-site energy distribution forms a structure quite distinct from the ideal case, as shown, as an example, in Fig. \ref{robust} for $N=6$ and $p=3$.
\begin{table}[h!]
\small
\centering
\begin{tabular}{|c|c|c|c|c|c|c|}
\hline
\rowcolor{gray!30}
$ N_{(p = 3)} $ & $ \textbf{4 s.f.} $ & $ \textbf{3 s.f.} $ & $ \textbf{2 s.f.} $ & $ \textbf{1 s.f.} $ & $ \textbf{5\% $\xi$} $ & $ \textbf{10\% $\xi$} $\\
\hline
4 & 99.9\% & 99.9\% & 99.8\% & 98.8\%& 99.9$ \pm $0.01\%&  99.8$ \pm $0.02\%\\
\hline
5 & 99.9\% & 99.9\% & 99.9\% & 97.7\% & 99.9$ \pm $0.01\%&  99.5$ \pm $0.03\%\\
\hline
6 & 99.8\% & 99.8\% & 99.4\% & 97.1\% & 99.7$ \pm $0.01\% & 99.3$ \pm 0.05$\%\\
\hline
7 & 96.7\% & 96.7\% & 95.6\% & 94.1\% & 96.5$ \pm $0.03\%&  95.9$ \pm $0.07\%\\
\hline
\end{tabular}
\begin{tabular}{|c|c|c|c|c|c|c|}
\hline
\rowcolor{gray!30}
$ N_{(p = 5)} $ & $ \textbf{4 s.f.} $ & $ \textbf{3 s.f.} $ & $ \textbf{2 s.f.} $ & $ \textbf{1 s.f.} $ & $ \textbf{5\% $\xi$} $ & $ \textbf{10\% $\xi$} $\\
\hline
4 & 96.1\% & 96.1\% & 96.0\% & 94.3\% & 95.9$ \pm $0.01\%&  95.8$ \pm $0.03\%\\
\hline
5 & 99.5\% & 99.5\% & 98.0\% & 90.3\%& 99.4$ \pm $0.02\%&  98.7$ \pm $0.06\%\\
\hline
6 & 97.8\% & 97.8\% & 97.4\% & 89.2\%& 97.5$ \pm $0.03\%&  96.6$ \pm $0.07\% \\
\hline
7 & 98.4\% & 98.4\% & 94.1\% & 87.9\% & 97.8$ \pm $0.04\%&  96.1$ \pm $0.11\%\\
\hline
\end{tabular}
\hfill
\begin{tabular}{|c|c|c|c|c|c|c|}
\hline
\rowcolor{gray!30}
$N_{(p = 7)}$ & $ \textbf{4 s.f.} $ & $ \textbf{3 s.f.} $ & $ \textbf{2 s.f.} $ & $ \textbf{1 s.f.} $ & $ \textbf{5\% $\xi$} $ & $ \textbf{10\% $\xi$} $\\
\hline
4 & 99.7\% & 99.6\% & 99.4\% & 98.1\% & 99.5$ \pm $0.01\%&  99.2$ \pm $0.03\%\\
\hline
5 & 98.7\% & 98.7\% & 92.2\% & 91.5\% & 98.4$ \pm $0.03\%&  97.9$ \pm $0.07\%\\
\hline
6 & 93.7\% & 93.6\% & 93.4\% & 83.6\% & 93.2$ \pm $0.04\%&  92.7$ \pm $0.1\%\\
\hline
7 & 89.4\% & 89.4\% & 89.3\% & 89.3\%& 89.0$ \pm $0.11\%&  87.6$ \pm $0.2\% \\
\hline
\end{tabular}
\caption{Transfer fidelity scores associated with the presented on-site energy configurations, with decreasing levels of precision, quantified by the reduction in the number of significant figures of, $\varepsilon_i$. This table demonstrates the robustness of the configurations shown for \( p = 3-7 \) within Fig. \ref{on}, for \( N \)-site chains ranging from 4 to 7 sites, evaluated up to 1 significant figure as a metric for the degree of experimental precision required. The last two columns, labeled `5\% $\xi$' and `10\% $\xi$', represent the mean fidelity obtained from 1,000 samples of asymmetric perturbations at 5\% and 10\% error strengths, respectively, illustrating the performance of the configurations under varying degrees of experimental noise.}
\label{table:efficiency}
\end{table}
The reduction in significant figures involves a different approach to that employed within \cite{Luke}, as it involves rounding numbers in the conventional sense, as opposed to a hard "cut" without arithmetic rounding. As an illustration, the value 1.452, containing four significant figures (4 s.f.), is rounded to 1.45 when expressed to three significant figures (3 s.f.) 1.5 rounded to two significant figures (2 s.f.) and finally 2 when reduced to one significant figure (1 s.f.). This approach provides another way to test the system's parameter error tolerance, offering further insight into the reproducibility of these protocols within potentially non-ideal settings. Furthermore, to evaluate the robustness of our protocols in more realistic environments, where errors are independent and thus mirror symmetry amongst the on-site energies is generally not maintained, we provide an analysis of error tolerance due to stochastic and asymmetric perturbations within Table.\ref{table:efficiency}. This analysis was conducted by defining an effective on-site energy 
\[
\varepsilon_i^{eff} = \varepsilon_i + J_{max} \, r_i \, \xi,
\]
where \( r_i \) is a randomly generated value within the range \([-0.5, 0.5]\), and \( \xi \) represents a predetermined error strength contribution (e.g. \( \xi = 0.1 \) for an error of 10\%). For the results presented within Table. \ref{table:efficiency} and Fig. \ref{Error}, the effective on-site energies are iterated over 1000 samples to collect the mean fidelity values and standard error. These findings demonstrate that asymmetric perturbations have stronger and more damaging effect on the results compared to the reduction in significant figures by one (from 4 to 3 as an example) alone, whilst strictly maintaining mirror symmetry. However, it is important to note that asymmetric errors in the range of 5-10\% are quite substantial and these results further underscore the robustness of these protocols to more severe environmental fluctuations/experimental, highlighting their capacity to maintain reliable transfer even under conditions of significant asymmetry and parameter variability.   

\section{Comparison with Previous Protocols}\label{prevprot}
\begin{figure}
    \begin{minipage}{.95\linewidth}
        \centering
        \includegraphics[width=1.0\linewidth]{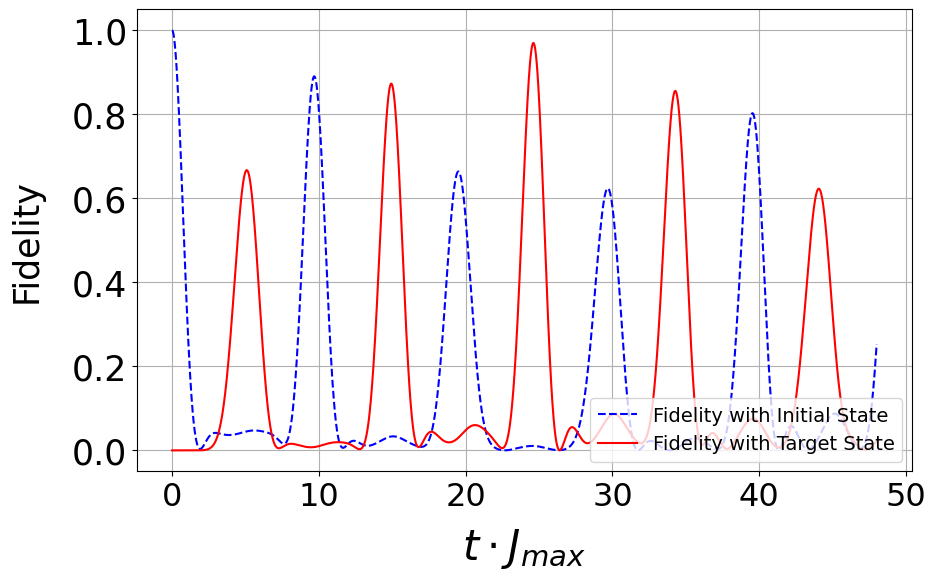} % Adjust the width value as needed
    \end{minipage}
    
    \vspace{0.5cm} % Adjust vertical spacing between figures if needed
    
    \begin{minipage}{0.95\linewidth}
        \centering
        \includegraphics[width=1.0\linewidth]{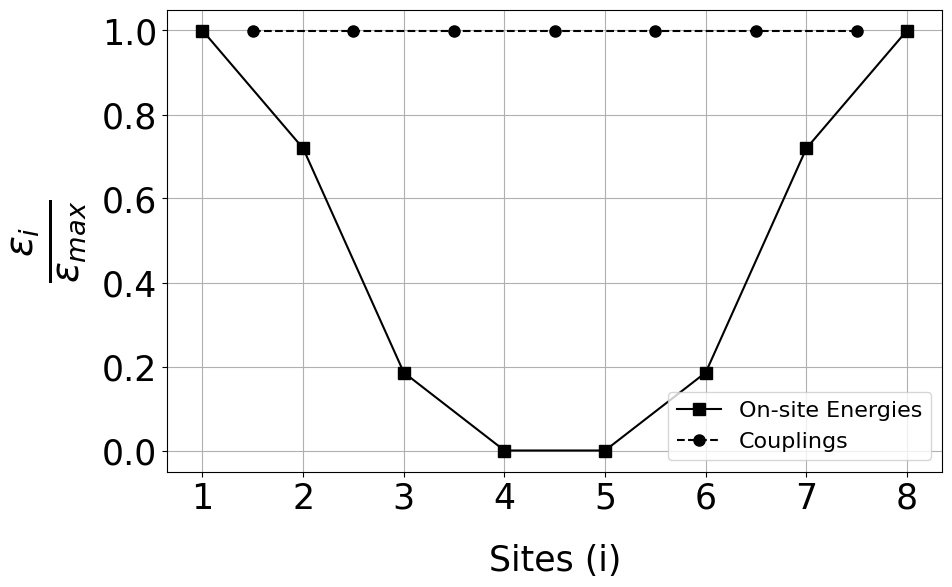} % Adjust the width value as needed
    \end{minipage}
    
    \caption{$N = 8$; $p = 5$ chain exhibiting high-fidelity state transfer (top) due to a anharmonic potential of the on-site energies (bottom). Highest fidelity attained within displayed time-window: 97.01\% at $t\cdot J_{max}$ = 24.61.}
    \label{N=8hf}
\end{figure}

Let us compare  findings from our approach with previous recommendations for the on-site energies to be site-dependent and parabolic \cite{G-B} 
\begin{equation}
    \varepsilon_i = \varepsilon_0|i - \frac{N}{2}|^2
    \label{B},
\end{equation}
where $\varepsilon_0$ is an initial strength. Using Eq.  (\ref{B}), our eigenvalue pinched solution for $N = 5$ ($p = 3$), corresponds to setting $\varepsilon_0 = 0.2930$ and may be conveniently reproduced in this manner to attain a fidelity of 99.98\% at $t \cdot J_{max} = 8.63$. Within \cite{G-B}, there was a recommended protocol for an 8-site chain, by setting $\varepsilon_0 = 0.5$, where high-fidelity ($\approx 98 \%$) was achieved after a time $t \cdot J_{max} \sim 10^4$, significantly longer than the timescale of results presented here. By employing an optimized solution from the genetic algorithm for \( N = 8 \), if we relax the strict structural requirement of the harmonic potential described by Eq. (\ref{B}), we can propose a solution for \( N = 8 \) with comparable fidelity, but with a transfer time orders of magnitude shorter. From Fig. \ref{N=8hf}, this configuration of the on-site energies (right hand panel), which clearly contains anharmonicity, can allow for QPST with fidelity \( 97\% \) at a time  $t \cdot J_{max} = 24.61$. This yields a significant speed up to the previous protocol ($t \cdot J_{max} = 10^4 \rightarrow 10$), with a marginal sacrifice in the fidelity. Larger fidelities might be possible at larger times, due to a different on-site energy configuration, but we have explored a maximum time window of $t \cdot J_{max} =  30$. As there is a clear incentive (discussed further in following section) to minimize the amount of time required for any $N$-site chain to transfer an initial encoding, we argue the homogeneously coupled configuration and dynamics displayed within Fig.   \ref{N=8hf} is more conducive to experimental proposal, provided that the anharmonic configuration can be implemented.

\section{Experimental Considerations}\label{sec:expcon}
The ability to tune on-site energies experimentally is crucial for implementing our proposals. It has been shown within linear ion traps/arrays of the sizes we have considered ($N \sim 10$), that simultaneous cooling of axial vibrational modes and non-uniform magnetic field gradients can achieve site-resolved Zeeman shifts, enabling individual control of the internal ionic resonances (shifting the resonances relative to each other) by a desired amount \cite{W3,W1,W2}. Furthermore, within cold atom arrays the use of site-varying external potentials are ubiquitous \cite{Bugnion2013,Paredes2004,Wieman1999}, which allows for the tuning of the on-site energies within hardware configured for high-fidelity single-spin addressing \cite{Weitenberg2011}. Similarly, for fabricated superconductor \cite{Caldwell,Zhao} or semiconductor qubits \cite{Botzem,Burkard}, it is usual to operate externally-controlled gates near to each qubit (to tune its on-site energy) and in between qubits (to tune coupling energies). The degree of difficulty associated to the task of effectively modulating on-site energies depends on the chosen hardware; however, as demonstrated within Sec. \ref{sec:results}, as long as the harmonic potential configuration among the on-site energies is approximately satisfied, high-fidelity state transfer can still be achieved. Since $J_{max}$ is hardware dependent and the times to achieve QPST occur within a time window of approx. $t \cdot J_{max} =  30$ (Figs. \ref{families} and \ref{N=8hf}), we may now produce a qualitative analysis observing how these protocols may be effectively incorporated within physical hardware. In reference to DiVincenzo's third requirement for a physically realizable quantum computer \cite{DiV}, we must ensure that \\
\begin{equation}
    \frac{T_2}{T_{QPST}} \gg 1,
    \label{Divv}
\end{equation}
\vspace{0.4cm}
and therefore that our time to QPST is much less than the time $T_{2}$ in which the system is likely to decohere. In defining $T_{QPST}$ as the time (in seconds) required for QPST, we need only to note the characteristic energy values intrinsic to candidate hardware environments \cite{Cheng,Xiang}.
\begin{table}[h!]
\centering
\small
\begin{tabular}{|l|c|c|c|}
\hline
\rowcolor{gray!30}
\hspace{1.cm}\textbf{Hardware} & \(\mathbf{{J_{max}/\hbar}}\) & \(\mathbf{T_{\text{QPST}}}\) &\(\mathbf{T_{2}}\) \\ \hline
Superconductor \cite{Xiang} & 10 GHz & 3 ns & 50 $\mu$s \\ \hline
Ion Trap \cite{Cheng}& 150 MHz & 0.2 $\mu$s & $\sim$ minutes \\ \hline
Semiconductor (Si-SiGe) \cite{Cheng}& 10 MHz & 3 $\mu$s &  50 $\mu$s \\ \hline
Neutral Atom Array \cite{Cheng}& 1 MHz & 30 $\mu$s & $\sim$ seconds \\ \hline
\end{tabular}
\caption{Comparison of different hardware in terms of operation time against decoherence. $J_{max}$ corresponds to the typical energy scale estimate derived from the two-qubit gate operation time $\tau_2$ described in the literature, as cited, and $T_{QPST}$ is the estimated time (in seconds) required for QPST (\(T_{QPST}= 30/J_{max})\). The Si-SiGe semiconductor was selected specifically as it allows for control over the largest number (to-date) of individually controlled semiconducting spin-qubits ($N = 6$).}
\label{tab:hardware}
\end{table}
Table.\ref{tab:hardware} demonstrates that our protocol fulfills the criteria (Eq. \ref{Divv})  sufficiently, allowing for a substantial number of operations/computational attempts before concerns related to decoherence become relevant. All, barring the semiconductor-based hardware, allow for several orders of magnitude of operations, within the frame of time considered. In reflecting that $T_{QPST}$ was taken simply as the largest time considered within Fig. \ref{families}, the specific times (in seconds) associated with the various number of sites ($p$-solutions) will be shorter, and hence the devices will be able to operate even longer. Then the number of attempts calculated from Eq. (\ref{Divv}) and the values in the last column of Table.\ref{tab:hardware} may be understood as a lower limit. This strongly indicates that our proposals could be utilized effectively within a variety of different physical hardware, for which the evolution of the system is governed by a Hamiltonian of the form Eq. (\ref{Hami}). 
\section{Conclusions} \label{sec:conclusions}
To conclude, we have demonstrated  through the use of a genetic algorithm, a set of novel solutions to yield QPST in uniformly coupled XY Hamiltonians. These are characterized by specific sets of configurations of on-site energies which agree with the analytical framework also discussed here. In addition, they offer an alternative spectrum structure to the previously known ones, corresponding to coupling configurations which yield PST. The general structure and approach of previous genetic algorithm investigations of spin chains/networks were maintained in this work, albeit with relevant modifications to the fitness function.  Future research endeavours  will focus on extending the optimization techniques employed herein to explore larger $N$-site chains. Furthermore, there is scope to investigate more of the physical/mathematical foundations behind the observed configuration patterns, particularly in understanding the parabolic orientation evident in configurations with $p > 1$. 
 Efforts are currently being made to explore these questions, though a detailed discussion lies beyond the scope of this paper. This work deepens our understanding of the dynamics of spin chains for quantum state transfer, and provides proposals which could be experimentally implemented within current state-of-the-art QIP structures.
\appendix*
\section{A} \label{sec:appendix}
\subsection*{Generalized Spectrum for \(N\)-Site Chain with Pinching by odd integer (\( p\))}

For an \(N\)-site chain, the eigenvalues \(E_n\), indexed from \(E_1\) to \(E_N\), (in increasing order) are given by:
\begin{equation*}
 E_n = \left\{
\begin{array}{ll}
\alpha((1-N) + 2(n-1)), & \text{for}\quad n = 1, 2, \ldots, N-1, \\
E_N = E_{N-1} + \frac{2\alpha}{p}, & \text{for } \quad n = N.
\end{array} \right.
\end{equation*}
where we have enforced a spectral spacing of $2\alpha$ and symmetry about 0 for convenience. It can be shown in general through explicit calculation that PST may be achieved, for arbitrary length $N$, at time $t_m = \frac{p\pi}{2\alpha}$. This spacing will always satisfy the condition Eq. (\ref{eig}), which is a finding from prior research on the necessary, systematic criteria for PST \cite{Kay}. Note, when $p = 1$, the system will mirror invert as a completely equidistant spectrum consistent with \cite{Christandl1}.
\subsection*{Equally-spaced spectra (p = 1)}
Considering an equally spaced spectrum for an N = 4 spin chain, such that
\[
E_i \in \left\{ 3\alpha, \alpha, -\alpha, -3\alpha \right\}
,
\]
where the spacing between sequential eigenvalues is $2\alpha$, and the eigenstates alternate in symmetry from the highest eigenenergy (even) to the lowest (odd). By applying time evolution to the even and odd symmetry eigenstates (with $\hbar = 1$), we can deduce a time in which the system has mirrored to itself, as

\[
|\psi(t)\rangle_+ = a_{3\alpha} e^{-i3\alpha t} |\phi_{3\alpha}\rangle + a_{-\alpha} e^{i\alpha t} |\phi_{-\alpha}\rangle,
\]
and
\[
|\psi(t)\rangle_- = a_{\alpha} e^{-i\alpha t} |\phi_{\alpha}\rangle + a_{-3\alpha} e^{3i\alpha t} |\phi_{-3\alpha}\rangle.
\]
Now, by setting \( t = \frac{\pi}{2\alpha} \), it follows that the eigenstates evolve to:
\[
|\psi\left(\frac{\pi}{2\alpha}\right)\rangle_+ = a_{3\alpha} e^{-i\frac{3\pi}{2}} |\phi_{3\alpha}\rangle + a_{-\alpha} e^{i\frac{\pi}{2}} |\phi_{-\alpha}\rangle,
\]
\[= e^{i\frac{\pi}{2}}(a_{3\alpha} e^{-i2\pi} |\phi_{3\alpha}\rangle + a_{-\alpha}|\phi_{-\alpha}\rangle)
\]
\[= e^{i\frac{\pi}{2}}|\psi(0)\rangle_+,
\]
and
\[
|\psi\left(\frac{\pi}{2\alpha}\right)\rangle_- = a_{\alpha} e^{-i\alpha t} |\phi_{\alpha}\rangle + a_{-3\alpha} e^{3i\alpha t} |\phi_{-3\alpha}\rangle.
\]
\[= e^{i\frac{\pi}{2}} (a_{\alpha}e^{-i\pi}|\phi_{\alpha}\rangle + a_{-3\alpha} e^{i\pi} |\phi_{-3\alpha}\rangle)
\]
\[= -e^{i\frac{\pi}{2}}|\psi(0)\rangle_-.
\]

Thus, through this demonstration, it can be observed that PST is guaranteed at time ($t = \frac{\pi}{2\alpha}$), for a particular configuration which yields a equally spaced energy spectrum, as our even and odd eigenstates acquire the same global phase, $e^{i\frac{\pi}{2}}$, with a relative minus sign for the odd states.\\
\subsection*{Pinch spectra (p = 3)}
\vspace{-0.8cm}
\[
E_n \in \left\{ 5\alpha/3, \alpha, -\alpha, -3\alpha \right\},
\]
The even (\(+\)) and odd (\(-\)) time-dependent wave functions are expressed as follows:
\[
|\psi(t)\rangle_+ = a_{5/3\alpha} e^{-i5\alpha t/3} |\phi_{5/3\alpha}\rangle + a_{-\alpha} e^{i\alpha t} |\phi_{-\alpha}\rangle,
\]
and
\[
|\psi(t)\rangle_- = a_{\alpha} e^{-i\alpha t} |\phi_{\alpha}\rangle + a_{-3\alpha} e^{3i\alpha t} |\phi_{-3\alpha}\rangle.
\]
Now, to demonstrate that such a system can exhibit perfect state transfer, consider substituting \( t = \frac{3\pi}{2\alpha} \). Upon applying the spectrum, the wave function solutions become:
\[
|\psi\left(\frac{3\pi}{2\alpha}\right)\rangle_+ = e^{i\frac{3\pi}{2}} \left( a_{5/3\alpha} |\phi_{5/3\alpha}\rangle + a_{-\alpha} |\phi_{-\alpha}\rangle \right),
\]
and
\[
|\psi\left(\frac{3\pi}{2\alpha}\right)\rangle_- = -e^{i\frac{3\pi}{2}} \left( a_{\alpha} |\phi_{\alpha}\rangle + a_{-3\alpha} |\phi_{-3\alpha}\rangle \right).
\]
The mirroring occurs because the odd states acquire a negative sign relative to the even states, resulting in the overall state mirroring about the center of the chain.

\end{document}